# Deep SR-ITM: Joint Learning of Super-Resolution and Inverse Tone-Mapping for 4K UHD HDR Applications

Soo Ye Kim      Jihyong Oh      Munchurl Kim

Korea Advanced Institute of Science and Technology

Daejeon, Republic of Korea

{sooyekim, jhoh94, mkimee} @kaist.ac.kr

## Abstract

*Recent modern displays are now able to render high dynamic range (HDR), high resolution (HR) videos of up to 8K UHD (Ultra High Definition). Consequently, UHD HDR broadcasting and streaming have emerged as high quality premium services. However, due to the lack of original UHD HDR video content, appropriate conversion technologies are urgently needed to transform the legacy low resolution (LR) standard dynamic range (SDR) videos into UHD HDR versions. In this paper, we propose a joint super-resolution (SR) and inverse tone-mapping (ITM) framework, called Deep SR-ITM, which learns the direct mapping from LR SDR video to their HR HDR version. Joint SR and ITM is an intricate task, where high frequency details must be restored for SR, jointly with the local contrast, for ITM. Our network is able to restore fine details by decomposing the input image and focusing on the separate base (low frequency) and detail (high frequency) layers. Moreover, the proposed modulation blocks apply location-variant operations to enhance local contrast. The Deep SR-ITM shows good subjective quality with increased contrast and details, outperforming the previous joint SR-ITM method.*

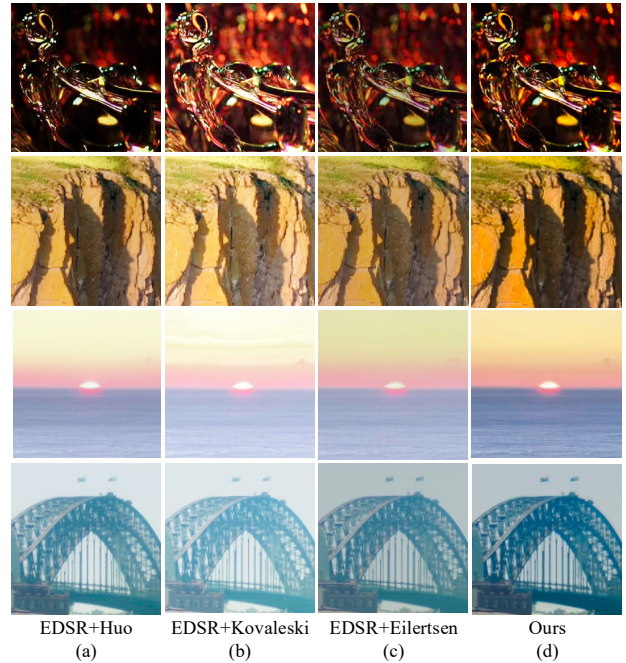

| EDSR+Huo | EDSR+Kovaleski | EDSR+Eilertsen | Ours |
|:---:|:---:|:---:|:---:|
| (a) | (b) | (c) | (d) |

Figure 1: Qualitative comparison with other methods. Our method shows enhanced contrast compared to the cascade of existing SR [20] and ITM methods [9, 13, 15].

## 1. Introduction

Modern TVs come with 4K/8K UHD (Ultra High Definition) displays and high dynamic range (HDR) capabilities. Nevertheless, the current Digital TV and Internet TV (IPTV) services still provide the legacy video contents of Full HD (FHD) resolution and standard dynamic range (SDR), which then, must be rendered on the premium TV displays that support 4K/8K UHD and HDR videos. Therefore at the terminal end, it is necessary to convert the FHD SDR videos to 4K/8K UHD HDR in order to display them on the premium displays. Furthermore, the new media services of high quality suffer from the lack of original 4K/8K UHD and HDR visual content. Thus, it is also essential to convert the legacy contents of Full HD and SDR video to 4K/8K UHD and HDR videos at the content production end.

In this paper, we aim to tackle the joint super-resolution (SR) and inverse tone-mapping (ITM) problem, where low-resolution (LR) SDR video can be directly converted into high–resolution (HR) HDR video. Along with the benefits in the terminal and content production ends, less bandwidth is necessary if the transmitted LR SDR videos are directly reconstructed as HR HDR on the device, with this joint SR-ITM framework. Moreover, users can benefit from high quality HR HDR visual content on their high-end TVs.

However, such joint SR-ITM is a complex problem. In LR images, high frequency details are lost with the reduced spatial resolution compared to HR images. In SDR images, local variations of contrast and local details are lost with the reduced signal range (amplitude) compared to HDR images. Therefore, for the joint SR-ITM task, it is important to jointly restore the fine details and contrast while increasing the spatial resolution and the signal amplitudes when predicting the HR HDR image from the LR SDR input.

In our proposed architecture, called Deep SR-ITM, the input image signal is decomposed into the base and detail layers, and separate feature extraction passes are designed for the two layers. This allows the network to focus on restoring details in the detail layer pass. For enhancing local contrast, convolution operations are not suitable as they are spatially equivariant (identical filters are applied across all pixels, in a sliding window manner). Thus, we design modulation blocks, which perform spatially-variant (pixel-location-specific) multiplication operations to modulate the local intensities. The produced modulation maps are also image-specific as they are produced image-dependently, unlike convolution filters that are fixed for all images once trained.

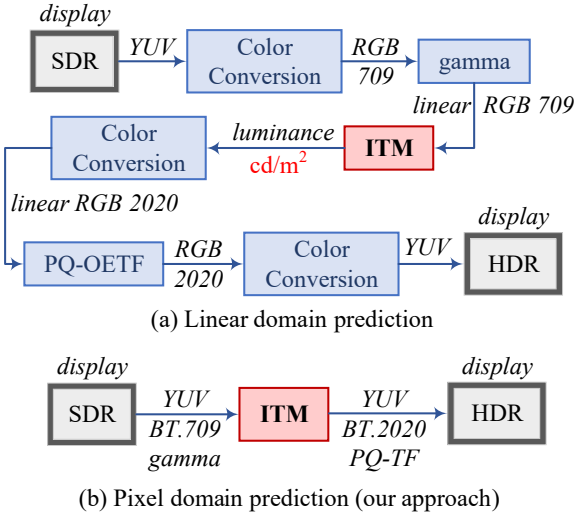

Figure 2: Comparison of the processing pipeline for rendering SDR to HDR video format between (a) linear domain, and (b) pixel domain prediction methods.

Our problem is functionally different from the previous ITM methods that aim to predict the luminance of the image in the linear domain, which is the physical brightness of the scene, typically in $cd/m^2$ (*candela per square meter*), as our network directly predicts HR HDR images in the HDR *display format*, in the pixel domain. Hence, the color gamut must be expanded from BT.709 to BT.2020 [2], the bit-depth increased from 8 bits/pixel to 10 bits/pixel , and the transfer function also changes from gamma [1] to PQ [3] or HLG [4] OETF. Fig. 2 compares the conventional luminance-predicting-ITM, and our approach that directly produces HDR videos in the pixel domain, when producing HDR images in display format. To facilitate real-world applications, we train and test our network with 4K (3,840×2,160) HDR videos.

Our contributions are three-fold:

- We introduce a novel deep network with modulation blocks that focus on enhancing local contrast for the joint SR-ITM problem.
- We incorporate input decomposition methods for the Deep SR-ITM to focus on the distinct low and high frequency components of the input image.
- For practicality, we experiment with 4K HDR videos to target realistic applications, and our network directly predicts HR HDR images in the HDR standard display format.

## 2. Related work

### 2.1. Inverse tone-mapping

Traditional single exposure ITM methods [5-13] exploit internal image characteristics, often focusing on the over-exposed (saturated) pixels [7, 10-12] in predicting the *luminance* of the scene. However, these methods concentrate on expanding the dynamic range and neglect the reconstruction of lost details and contrast. The recent data-driven approaches in single exposure ITM [14, 15] indirectly resolved this issue through large amount of training data. Zhang *et al*. [14] focused on outdoor images and predicted the degree of sun elevation simultaneously, and Eilertsen *et al*. [15] trained a U-Net [41] structure only for the saturated regions and later blended the prediction with the input SDR image for the unsaturated regions.

However, for our problem where color changes occur for all regions (saturated and unsaturated), it is unsuitable to convert only the saturated regions. The method in [14] cannot be applied for our problem since it only targets images with the sun in the sky. Furthermore, the previous methods predict the luminance (in $cd/m^2$) and therefore, do not consider the color gamut expansion. Architecture-wise, the methods in [14, 15] employ auto-encoder-type structures (e.g. U-Net). However, for joint SR-ITM, relying fully on auto-encoder structures may result in losing important spatial-wise information that is crucial for reconstructing higher resolution images for the SR side. Therefore, we specifically design our network to focus on restoring the lost details and enhancing local contrast, for all pixel regions. The Deep SR-ITM, trained on 4K HDR videos, directly predicts the HR HDR image in the HDR display format for practical applications on HDR TVs.

### 2.2. Super-resolution

Starting with Dong *et al*.'s method [16], many CNN-based SR methods [17-24] have been proposed. After the pixel shuffler method [17] and the very deep SR with residual learning [18], deep networks have become more and more complex, with more recent structures employing residual blocks [19-20, 22-23], dense connections [21-22] and channel attention blocks [23]. Our method is highly inspired by very deep networks that are able to generate finer details. We employ different combinations of residual, skip and modulation blocks, for reconstructing space-wide high frequency details, and restoring amplitude-wise local contrast. We let our network focus on the separate aspects by decomposing the input using the guided filter [26].

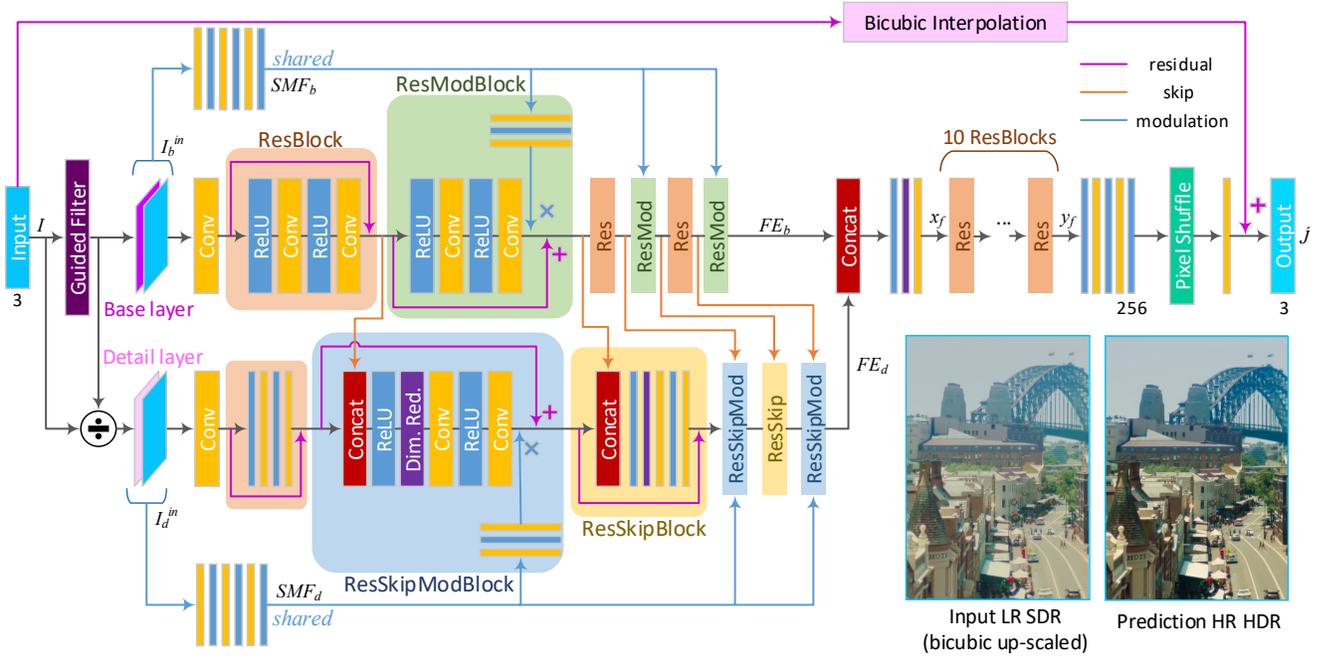

Figure 3: Network architecture of our Deep SR-ITM with ResBlock, ResModBlock, ResSkipBlock and ResSkipModBlock. The input LR-SDR image (*aqua blue* box) is concatenated with its guided filter decompositions (base and detail layer) before entering the network.

### 2.3. Joint SR-ITM

A natural way to generate HR HDR images from LR SDR ones is to cascade existing SR and ITM methods in series. However, this is inefficient, as it increases overall complexity, and inaccurate, as cascading may accumulate the error made from the previous prediction [25]. An end-to-end jointly trainable SR-ITM network was first proposed by Kim *et al.* [25], where a multi-purpose CNN structure was designed to simultaneously perform three tasks of SR, ITM and joint SR-ITM in a single network with three output branches. The multi-purpose CNN simply consists of stacked convolution layers for the SR and ITM branches, and a third joint SR-ITM branch that makes use of the concatenated feature maps of the other two branches. On the other hand, the proposed Deep SR-ITM tackles the joint SR-ITM problem in a more elaborate and dedicated way by using the decomposed signal components of base and detail information for spatially-variant modulations.

## 3. Proposed method

We propose Deep SR-ITM, a deep residual network based on signal decomposition and modulations, where an HR HDR image in the HDR display format of BT.2020 [2] and PQ-OETF [3] is generated from a single LR SDR image. Our network architecture is shown in Fig. 3.

### 3.1. Input decomposition

Before entering the network, the input LR SDR image $I$ is decomposed into the base layer $I_b$ and the detail layer $I_d$ using the guided filter (an edge-preserving low-pass filter) [26]. $I_b$ is computed by applying the guided filter to $I$, and then $I_d$ is obtained by simply dividing $I$ by $I_b$ as

$$I_d = I \oslash I_b, \tag{1}$$

where $\oslash$ denotes element-wise division. $I_b$ contains a blurred *color* image, dominant with low frequency information, and $I_d$ is mostly *colorless*, dominant with high frequency information (e.g. edges and texture).

Since $I$ also contains useful information, it is concatenated along with $I_b$ and $I_d$ in the channel direction.

$$I_b^{in} = [I \ I_b], \quad \text{and} \quad I_d^{in} = [I \ I_d]. \tag{2}$$

Then, $I_b^{in}$ and $I_d^{in}$ proceed separately in two distinct feature extraction passes, so that the top base layer pass can concentrate on converting the color and expanding the amplitude, and the bottom detail layer pass can focus on restoring high frequency details.

### 3.2. Residual skip modulation blocks

**Modulation.** Convolution operations in convolution layers are spatially equivariant, since the same convolution filters are applied to all pixel positions. Especially for enhancing local contrast, this property of convolutions limits the capability of the network, as image characteristics (such as the contrast to be restored) vary depending on pixel locations. Furthermore, convolution filters are fixed once the network is trained, and the same filters are applied to all image samples. Therefore, we introduce spatially-variant and image-adaptive modulations by element-wise multiplication, to aid the network in modelling more complex mappings, than can be modelled by simple CNNs. Operation-wise, this is similar to attention blocks (actually, a generalization of spatial channel attention) in high level vision tasks, such as object detection and classification. For those tasks, attention blocks help the network focus on semantically important regions. For a low level vision task like joint SR-ITM, the location-specific multiplication operation helps *modulate* the image signal pixel-by-pixel.

**Residual blocks.** We design four different combinations of residual, skip and modulation blocks: ResBlock, ResModBlock, ResSkipBlock and ResSkipModBlock, in our network in Fig. 3.

Firstly, if we let $x$ be an input to the $i$-th block, the output of the $i$-th ResBlock $RB_i$ (*orange* box) can be expressed as,

$$RB_i(x) = (Conv \circ RL \circ Conv \circ RL)(x) + x = C_{RB}(x) + x, \quad (3)$$

where *Conv* is a convolution layer and *RL* is the ReLU activation [27] ($RL(\cdot) = \max(0, \cdot)$).

Secondly, the ResModBlock (*green* box) has an additional modulation component. It requires the shared modulation features ($SMF_b$) of the base layer given by,

$$SMF_b = (RL \circ Conv \circ RL \circ Conv \circ RL \circ Conv)(I_b^{in}). \quad (4)$$

The modulation component then goes through additional layers that are *not shared* with other ResModBlocks, to account for the difference depending on the depth of each block. The output of the $i$-th ResModBlock $RMB_i$ is then given by,

$$RMB_i(x) = C_{RB}(x) \odot \{(Conv \circ RL \circ Conv)(SMF_b)\} + x, \quad (5)$$

where $\odot$ denotes element-wise multiplication.

The output ($FE_b$) of the last feature extraction layer of the top base layer pass is then given by,

$$FE_b = (RMB_m \circ RB_m^b \circ ... \circ RMB_1 \circ RB_1^b \circ Conv)(I_b^{in}), \quad (6)$$

where $FE_b$ is obtained by alternatively applying the ResBlock and the ResModBlock, and contains $m$ $RB^b$s (ResBlocks in the base layer pass) and $m$ $RMB$s.

For the detail layer pass, skip components are additionally used to aid the flow of information. The third type of block, ResSkipBlock (*yellow* box), bridges the features of the ResModBlock in the base layer pass. The output of the $i$-th ResSkipBlock $RSB_i$ is given by,

$$RSB_i(x) = (Conv \circ RL \circ Conv \circ DR \circ RL)([x \; RMB_i]) + x, \quad (7)$$

where *DR* is a dimension reduction layer with 1×1 convolutions, and $[x \; y]$ denotes the concatenation of x and y in the channel direction, as in Eq. (2). The DR layer acts as a selection module that controls which and how much information to pass through from the expanded input.

Lastly, the ResSkipModBlock is designed with modulations as well as skip connections. The output of the $i$-th ResSkipModBlock $RSMB_i$ is given by,

$$RSMB_i(x) = \{(Conv \circ RL \circ Conv \circ DR \circ RL)([x \; RB_i^b])\} \odot \\ \{(Conv \circ RL \circ Conv)(SMF_d)\} + x. \quad (8)$$

The output ($FE_d$) of the last feature extraction layer of the bottom detail layer pass is then given by,

$$FE_d = (RSMB_m \circ RSB_{m-1} \circ ... \circ RSB_1 \circ RSMB_1 \circ RB_1^d \circ Conv)(I_d^{in}). \quad (9)$$

$FE_d$ contains $m$ $RSMB$s, $m$-1 $RSB$s, and 1 $RB^d$ (ResBlock in the detail layer pass).

### 3.3. Fusion and synthesis

The later parts of the Deep SR-ITM consist of fusing the features of the base layer and the detail layer ($FE_b$ and $FE_d$), and finally producing the HR HDR output. ResBlocks are again used for the integration part, denoted as $RB^f$. The input to the ResBlock, $x_f$, is given by,

$$x_f = (Conv \circ DR \circ RL)([FE_b \; FE_d]), \quad (10)$$

and the output $y_f$ after the $n$-th ResBlock is expressed as,

$$y_f = (RB_n^f \circ ... \circ RB_1^f)(x_f). \quad (11)$$

Then, the final HR HDR prediction $\hat{J}$ is given by,

$$\hat{J} = (Conv \circ PS \circ RL \circ Conv \circ RL \circ Conv \circ RL)(y_f) + Bic(I), \quad (12)$$

where *PS* denotes pixel shuffle [17], and $Bic(I)$ denotes bicubic up-scaling of $I$ to match the resolution of $\hat{J}$. Global residual learning is applied, as well as the local residual learning inside the four types of residual blocks to ease training and enhance the prediction accuracy.

### 3.4. Toy network

We design a separate toy network (simplified version of the Deep SR-ITM) to further analyze the effect of input decompositions (Sec. 4.2.) and evaluate different types of modulations (Sec. 4.3.). The toy network should be simple for the efficient management of experiments, but also representative of the original Deep SR-ITM so that experiment results on the toy network mirror those of the original Deep SR-ITM. The toy network is illustrated in Fig. 4. It contains global and local residual connections, skip connections from the base to detail layer pass, and modulations for each pass, as well as input decomposition.

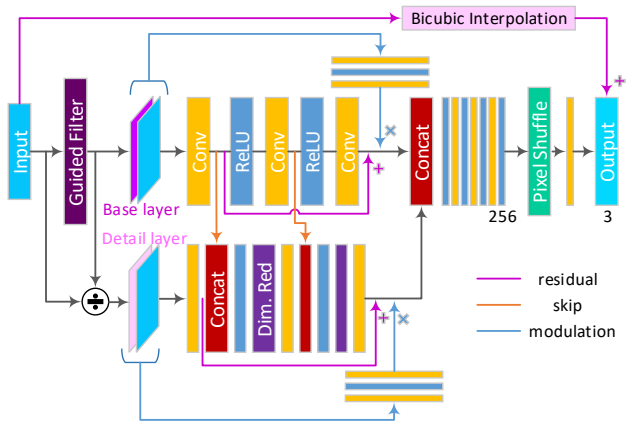

Figure 4: Network architecture of the toy network.

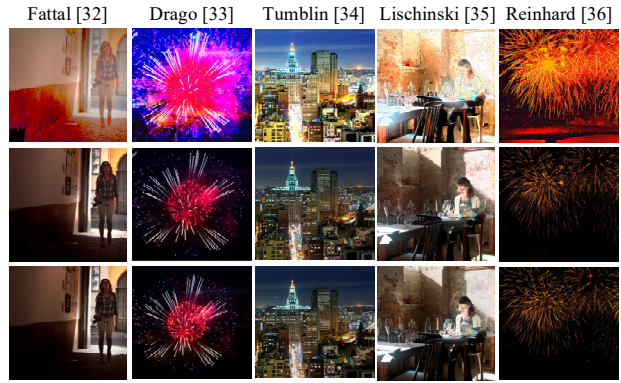

Figure 5: Tone-mapping examples. References at the top indicate the TMO used for producing images in the top row. On the second row are SDR images produced from YouTube, and on the last row are the original HDR images.

## 4. Experiment results

### 4.1. Experiment conditions

**Implementation details.** All convolution filters are of size 3×3 with 64 output channels, except for the layer before pixel shuffle with 256 channels, and the output layer with 3 channels. The network is trained and tested for all three YUV channels as the color container is also converted. In the network architecture, $m = 3$ and $n = 10$. For experiments with the scale factor (SF) of 4 for SR, two pixel shufflers were implemented, with a convolution layer in between them. All results with the SF unmentioned are for SF = 2.

**Data.** We collected ten 4K-UHD HDR videos (BT.2020 with PQ-OETF) of 59,818 frames in total from YouTube. Among these, 7 videos were used for training (44K frames) and 3 were left for testing. 20 to 40 patches of size 160×160 were randomly cropped within each frame. To avoid high coherence among the frames, patches were extracted with a frame stride ranging from 10 to 80. For testing, we selected 28 different scenes in the test video. For tone-mapping to obtain SDR videos from their corresponding HDR ones, we investigated using 19 different TMOs selected from the implementations in HDR Toolbox [28]. In this case, the HDR videos were linearized and the color container was converted (BT. 2020 to BT. 709) prior to tone-mapping, and gamma-encoded afterwards. However, these methods exhibited unnatural colors, and produced unstable results for dark scenes as shown in Fig. 5. Since the LR SDR data should look natural and be as close to real SDR videos as possible, we gathered the SDR video pairs through the automatic conversion process of YouTube, and down-scaled the frames with bicubic filtering.

**Training.** We used the L2 loss, Adam [29] optimizer and Xavier initialization method [30] for training. L2 loss works well, as the HR HDR prediction is in the pixel domain. The Deep SR-ITM is pre-trained without modulation (without the *blue* arrows in Fig. 3) for 490K iterations, with the learning rate of $5\times10^{-7}$ for weights and $5\times10^{-8}$ for biases. After pre-training, the network was fully trained *with* modulation for another 660K iterations. We take this two-step training strategy so that the modulation maps are able to train on meaningful feature maps. The mini-batch size is 16. The whole training and testing processes were implemented using MatConvNet [31].

For the toy network, we used separately collected SDR-HDR pairs. It is also pre-trained first without modulation (without the *blue* arrows in Fig. 4), and is fully trained afterwards. The learning rate is set to $10^{-6}$, with other training parameters set the same as the Deep SR-ITM.

**Visualization.** All visualizations of the HDR results in this paper are obtained through the madVR renderer with the MPC-HC player.

### 4.2. Input decomposition

We first analyzed the effect of input decomposition, by designing five variations of the toy network, with *one* to *three* feature extraction passes, each with possible combinations of the LR SDR image, its base layer and its detail layer as input. The network designs are summarized in Table 1. In all cases, no modulations were involved.

Specifically the network in column (a) of Table 1, is a single-pass network with only the image (without the base nor the detail layer) entered as input. For the network in column (b), all three layers were stacked before entering the single-pass network. Column (c) describes a two-pass network as in Fig. 4 but *without* stacking the image. The network in column (d) is identical to the network in Fig. 4, where each of the base and detail layers are stacked with the LR SDR image prior to entering the network. The network in column (e) has three feature extraction passes, each for the image, base and detail layer. The number of filter parameters is matched for all networks in (a)-(e).

|  | (a) | (b) | (c) | (d) | (e) |
|---|---|---|---|---|---|
| pass | 1 | 1 | 2 | 2 | 3 |
| image | ✓ | ✓ | ✗ | ✓ | ✓ |
| base/detail | ✗ | ✓ | ✓ | ✓ | ✓ |
| stack | ✗ | ✓ | ✗ | ✓ | ✗ |
| PSNR(dB) | 38.11 | 38.25 | 38.11 | **38.46** | 38.21 |
| SSIM | 0.9905 | 0.9907 | 0.9911 | **0.9916** | 0.9914 |

Table 1: Effect of input decompositions and concatenations.

The following can be analyzed from Table 1:

(i) For single-pass structures, decomposing the input and *stacking* all three layers, as in column (b), results in 0.14 dB increase in PSNR (compared to column (a)).
(ii) For two-pass structures, stacking (column (d)) brings 0.35 dB PSNR gain, with decomposed layers.
(iii) Designing *separate passes* for the base and detail layers (column (d)) is important with 0.21 dB PSNR gain over column (b).
(iv) The undecomposed image is more useful when *stacked*, by comparing columns (b), (d) and (e).

Most importantly, designing separate passes for the base and detail layers, and stacking the LR SDR image to each pass as a guidance as in column (d) obtains the highest gain in performance over the baseline of column (a).

### 4.3. Modulation

**Input combinations.** We explored using different combinations of using the image, base and detail layers to extract the shared modulation features (*SMF*) as shown in Table 2. Stacking decomposed layers for modulations results in maximum 0.08 dB PSNR gain in Table 2, which is less effective than for the network input (0.35 dB PSNR gain in Table 1). This implies that specific features, such as the overall brightness in the base layer or the high frequency components in the detail layer, are sufficient for modulation, while the integrated image is necessary for the main pass in order to restore a full HR HDR image.

**Visualization of modulation maps.** We refer to the modulation features that are multiplied to the main branch feature maps at each modulation block as modulation maps. Fig. 6 shows the input image and the modulation maps of the base and detail layer pass. The modulation maps in Fig. 6 are taken from $RMB_1$ for the base layer pass, and $RSMB_1$

| Input to $SMF_b$ | image | base | [image base]* |
|---|---|---|---|
| Input to $SMF_d$ | image | detail | [image detail]* |
| PSNR (dB) | 38.43 | 38.44 | **38.52** |
| SSIM | 0.9918 | 0.9918 | **0.9920** |

*[x y] denotes the concatenation of x and y

Table 2: Effect of using various combinations of inputs for extracting the shared modulation features of the base ($SMF_b$) and detail layer ($SMF_d$) passes.

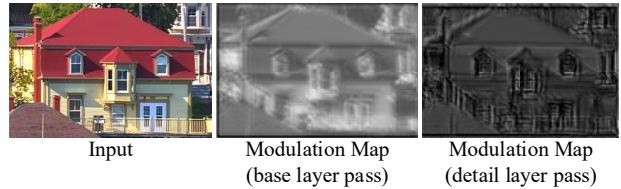

Figure 6: Modulation maps of base and detail layer passes.

for the detail layer pass. It can be verified that the modulations are being performed on the intensity for the base layer, and on edges and details for the detail layer.

### 4.4. Ablation study

We performed an ablation study on three main components of the Deep SR-ITM. Table 3 summarizes the results of this ablation study. The guided filter decomposition and modulation yield 0.08 dB and 0.22 dB PSNR performance gain consecutively, with a total of 0.3 dB gain compared to the baseline network (column (a)) without decomposition nor modulations. Also in Table 3, the skip connections are only effective without modulation. The benefits of skip connections and modulations seem to overlap, as they both help with the flow of information.

|  | (a) | (b) | (c) | (d) | (e) |
|---|---|---|---|---|---|
| GF* | ✗ | ✓ | ✓ | ✓ | ✓ |
| Skip | ✗ | ✗ | ✓ | ✗ | ✓ |
| Modulation | ✗ | ✗ | ✗ | ✓ | ✓ |
| PSNR (dB) | 35.29 | 35.37 | 35.44 | **35.59** | 35.58 |
| SSIM | 0.9730 | 0.9732 | 0.9734 | **0.9747** | 0.9746 |

*GF = Guided Filter

Table 3: Ablation study.

### 4.5. Performance comparisons

**Compared methods.** We compare against the cascade of a single image SR method [20] and ITM methods [9, 12, 13, 15], and also the joint SR-ITM method in [25]. We used the official implementation provided by the authors for methods in [15, 20], and the implementations in HDR Toolbox [28] for methods in [9, 12, 13]. For the ITM methods of [9, 12, 13], the SDR input was linearized prior to the ITM process. For all ITM methods in [9, 12, 13, 15], color was converted from RGB709 to RGB2020 [2], and PQ-OETF [3] was applied after ITM, following the post-ITM pipeline in Fig. 2-(a). The maximum brightness was set to 1,000 cd/m$^2$, following the HDR10 standard for HDR TVs. Note that methods in [9, 12, 13] are not data-driven methods (thus *not* dependent on training data), but *expansion operators*, applicable to any SDR image. Because Eilertsen *et al*'s method failed to train on our data (which do not contain sufficient saturated regions compared to their *HDR luminance* data in cd/m$^2$), we experimented with their released test code without re-training. For the joint SR-ITM method in [25], we re-trained their multi-purpose CNN architecture for SF = 2, and SF = 4 on the same training data as ours.

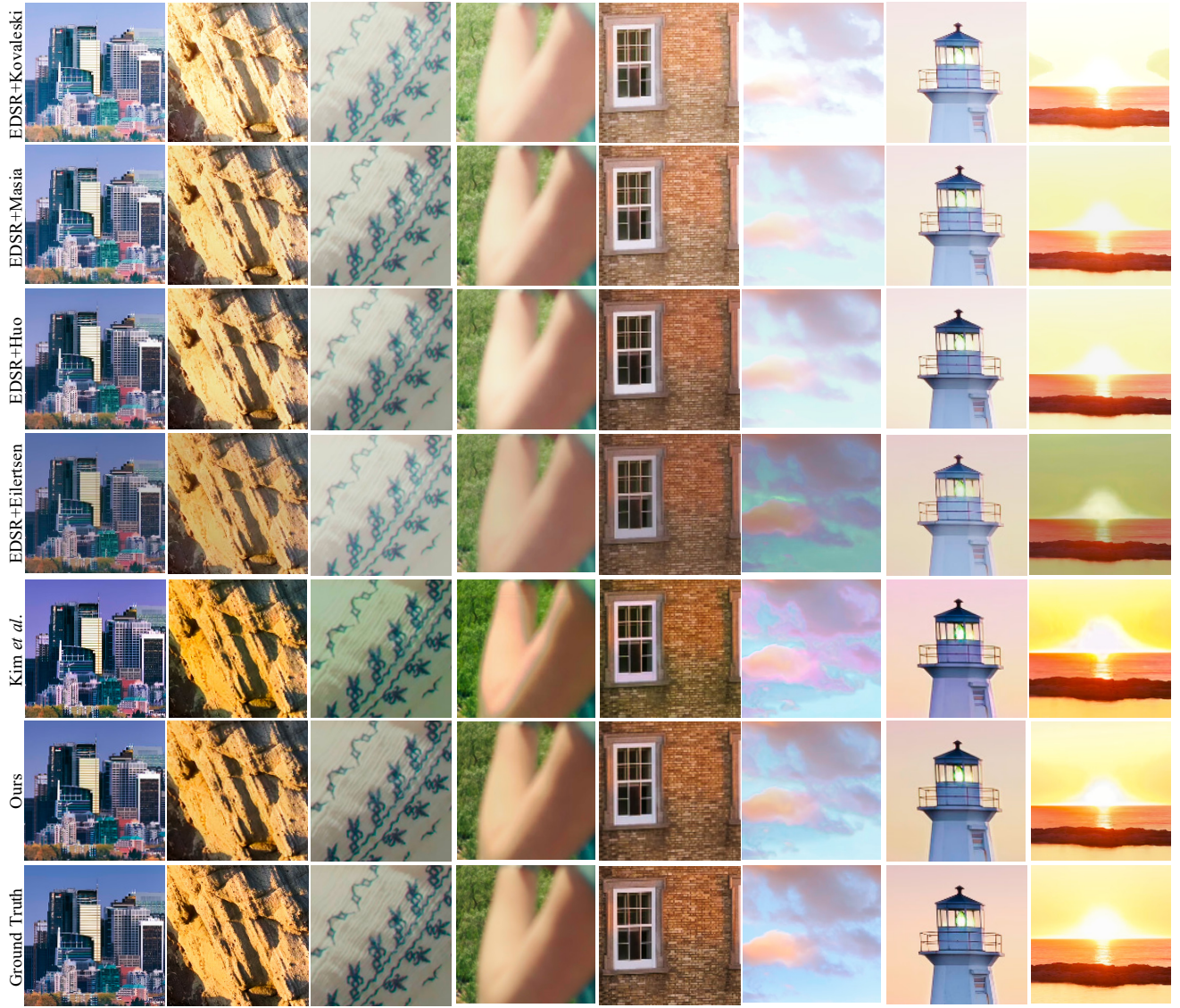

Figure 7: Qualitative comparison against EDSR [20] cascaded with ITM methods in [9, 12, 13, 15], and the joint SR-ITM method in [25].

**Qualitative comparison.** We provide the qualitative comparison in Fig. 1 and Fig. 7. In Fig. 7, the cascaded methods lack overall contrast, and the results by [25] shows good contrast, but the produced colors are unnatural with obvious artifacts in some cases. Our method demonstrates natural colors with enhanced contrast and restored details. More results are provided as the Supplementary Material.

**Quantitative comparison.** We provide the quantitative results on five metrics: PSNR, multi-exposure PSNR –mPSNR- [28], SSIM [37], multi-scale SSIM -MS-SSIM- [38] and HDR-VDP-2.2.1 [39]. For mPSNR, the 8-bit quantization in the pipeline is modified to 10-bit quantization, and PSNR is averaged for exposure values from -3 to +3. For HDR-VDP, the linearized Y channel was compared with the 'luminance' option and 'pixel per degrees' set to 30. The quantitative comparison on SF = 2 and SF = 4 is given in Table 4, where the number after ± denotes the standard deviation. Deep SR-ITM outperforms the previous methods in all measures and scale, except HDR-VDP for SF = 4.

**Visualization of feature maps.** Fig. 8. shows the three layers (image, base and detail) used as input and the intermediate feature maps (conv1, $FE_b$, $FE_d$ and $y_f$), where conv1 is produced after the first convolution layer of each pass in the main network. Features with edges and texture are extracted in the detail layer pass, while features with

| Method | Scale | PSNR (dB) | mPSNR (dB) | SSIM | MS-SSIM | HDR-VDP (Q) |
|---|---|---|---|---|---|---|
| EDSR [20]+Kovaleski *et al*. [9] | ×2 | 23.59 ± 0.95 | 25.48 ± 2.07 | 0.6504 ± 0.1545 | 0.9737 ± 0.0155 | 57.24 ± 3.62 |
| EDSR [20]+Masia *et al*. [12] | ×2 | 24.71 ± 3.21 | 26.65 ± 5.13 | 0.7095 ± 0.1158 | 0.9679 ± 0.0324 | 60.04 ± 5.68 |
| EDSR [20]+Huo *et al*. [13] | ×2 | 29.76 ± 2.62 | 31.81 ± 3.90 | 0.8934 ± 0.0717 | 0.9764 ± 0.0166 | 58.95 ± 6.41 |
| EDSR [20]+Eilertsen *et al*. [15] | ×2 | 25.80 ± 3.80 | 28.22 ± 4.39 | 0.7586 ± 0.1638 | 0.9635 ± 0.0204 | 53.51 ± 7.73 |
| Multi-purpose CNN [25] | ×2 | 34.11 ± 2.61 | 36.38 ± 3.70 | 0.9671 ± 0.0160 | 0.9817 ± 0.0101 | 60.91 ± 5.12 |
| **Deep SR-ITM (Ours)** | ×2 | **35.58** ± 4.80 | **37.80** ± 5.83 | **0.9746** ± 0.0133 | **0.9839** ± 0.0097 | **61.39** ± 5.82 |
| EDSR [20]+Kovaleski *et al*. [9] | ×4 | 23.46 ± 0.99 | 25.34 ± 2.09 | 0.6325 ± 0.1621 | 0.9670 ± 0.0150 | 56.70 ± 3.72 |
| EDSR [20]+Masia *et al*. [12] | ×4 | 24.54 ± 3.18 | 26.47 ± 5.10 | 0.6968 ± 0.1173 | 0.9608 ± 0.0318 | **57.74** ± 5.27 |
| EDSR [20]+Huo *et al*. [13] | ×4 | 28.90 ± 2.15 | 30.92 ± 3.42 | 0.8753 ± 0.0804 | 0.9693 ± 0.0156 | 55.59 ± 6.47 |
| EDSR [20]+Eilertsen *et al*. [15] | ×4 | 26.54 ± 2.69 | 28.75 ± 3.20 | 0.7822 ± 0.1951 | 0.9631 ± 0.0161 | 53.88 ± 5.79 |
| Multi-purpose CNN [25] | ×4 | 33.10 ± 3.58 | 35.26 ± 4.62 | 0.9499 ± 0.0230 | 0.9758 ± 0.0104 | 56.41 ± 6.03 |
| **Deep SR-ITM (Ours)** | ×4 | **33.61** ± 4.32 | **35.73** ± 5.24 | **0.9561** ± 0.0259 | **0.9748** ± 0.0109 | 56.07 ± 6.83 |

Table 4: Quantitative performance comparison.

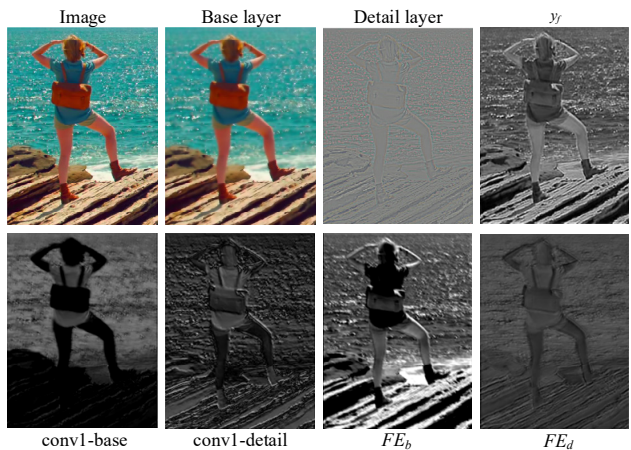

Figure 8: Visualization of intermediate feature maps.

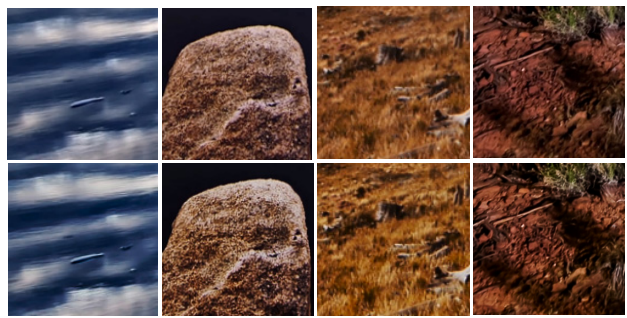

Figure 9: Prediction results for realistic conditions. On the top row are the LR SDR inputs up-scaled using bicubic interpolation, and on the bottom row are the HR HDR predictions.

overall brightness and global contrast are extracted from the base layer. Later layers are able to concentrate on such features. $y_f$, produced during the fusion stage, already seems like an integrated gray scale image

**Runtime evaluation.** Our method takes 5.85 seconds and 5.05 seconds to generate a 4K frame for SF = 2 and SF = 4, respectively, on an NVIDIA TITAN Xp GPU. The total number of the filter parameters of Deep SR-ITM is 2.5M for SF = 2, and 2.64M for SF = 4. Note that the number of filter parameters of EDSR alone amounts to 43M.

### 4.6. Reconstruction under realistic conditions

We test our Deep SR-ITM on various videos that are *originally* 4K SDR (not tone-mapped), to verify its robustness in realistic conditions. As ground truth HDR videos do not exist for these videos, we only provide the visual results in Fig. 9. The Deep SR-ITM is still able to generate the HR HDR reconstructions with enhanced contrast and details under more realistic conditions.

## 5. Conclusion

In this paper, we proposed the Deep SR-ITM, a joint SR-ITM framework, where the low and high frequency information in the input SDR images are decomposed. Thanks to this input decomposition strategy, Deep SR-ITM is able to precisely predict the lost high frequency details assisted by the detail layer for spatial up-scaling, while simultaneously expanding the overall intensity and color to the HDR brightness context assisted by the base layer, for the ITM task. A modulation scheme is incorporated to boost the local contrast in the image signal amplitudes by introducing spatially-variant operations. Directly producing HR HDR images in the pixel domain is a very handy application for generating premium visual content for UHD HDR consumer displays. All relevant codes and the information on the test set are available in GitHub.

**Acknowledgement.** This work was supported by Institute for Information & communications Technology Promotion (IITP) grant funded by the Korea government (MSIT) (No. 2017-0-00419, Intelligent High Realistic Visual Processing for Smart Broadcasting Media).

# Deep SR-ITM: Joint Learning of Super-Resolution and Inverse Tone-Mapping for 4K UHD HDR Applications

## -Supplementary Material-

## 1. Additional qualitative comparisons

### 1.1. Lack of contrast in the cascade of EDSR and Eilertsen *et al.*'s method

In Fig. 1, we provide more qualitative comparisons of our proposed method, Deep SR-ITM, against the cascade of EDSR [5], which is the state-of-the-art super-resolution (SR) method and Eilertsen *et al.*'s method [4], which is a recent CNN-based single exposure inverse tone-mapping (ITM) method. Our method shows enhanced contrast compared to the cascaded method of [5] and [4].

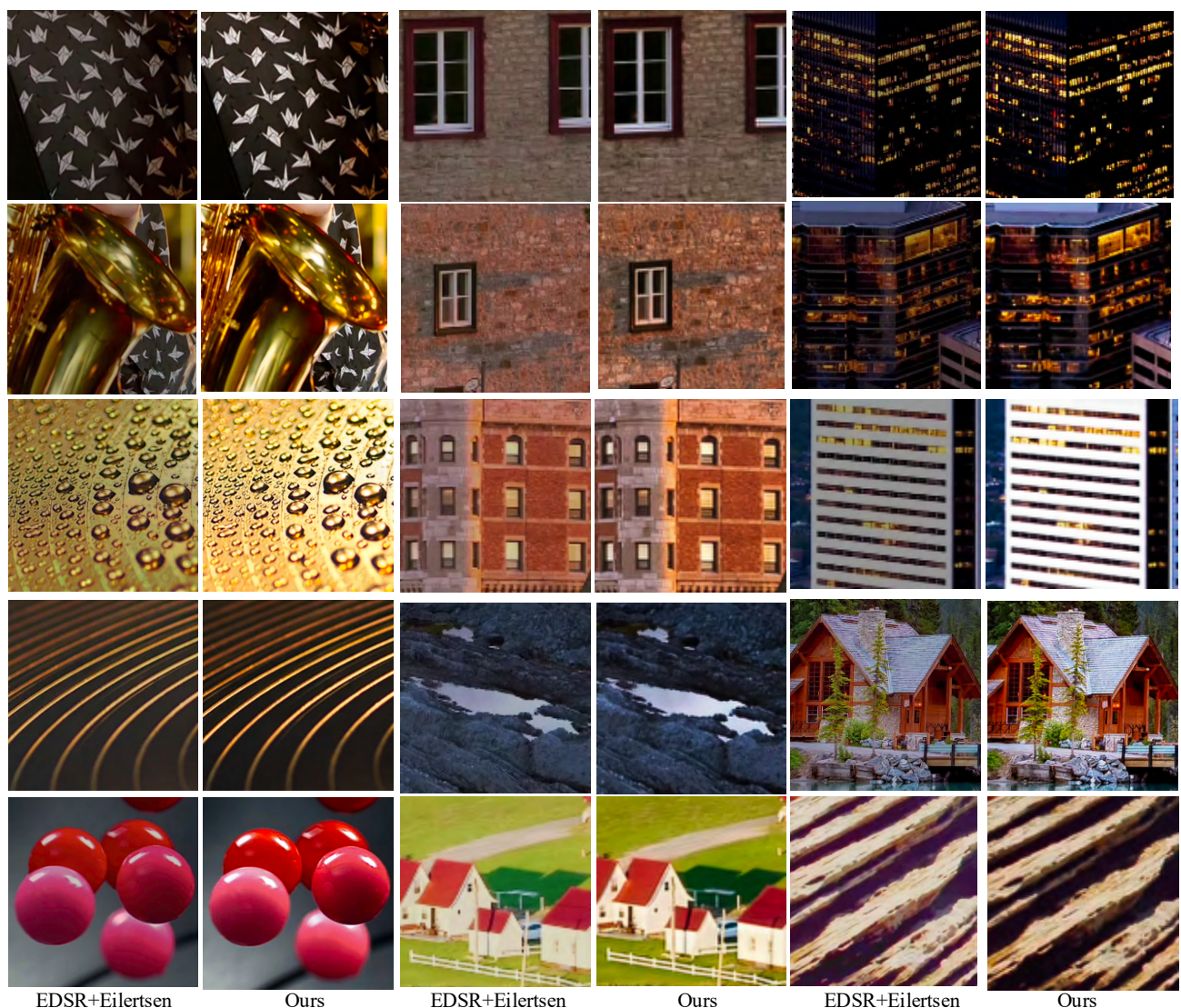

EDSR+Eilertsen    Ours    EDSR+Eilertsen    Ours    EDSR+Eilertsen    Ours

Figure 1: Qualitative comparison against the cascade of EDSR [5] and Eilertsen *et al.*'s method [4].

## 1.2. More comparisons against the cascade of conventional methods

In Fig. 2, we provide more qualitative comparisons of Deep SR-ITM against the cascade of EDSR [5] and ITM methods [1-4] compared in our original paper. We follow the same experiment protocol as presented in Sec. 4.5 in our original paper. Our method restores local contrast and details while expanding the signal amplitudes and increasing the overall resolution. In order to produce high resolution (HR) high dynamic range (HDR) videos in display format, these methods should be connected in cascade to perform the joint task, linearized before the ITM process, and the color container and the transfer function must be converted manually, which is cumbersome. In contrast, the proposed Deep SR-ITM directly generates HR HDR video frames in display format, with enhanced contrast and details.

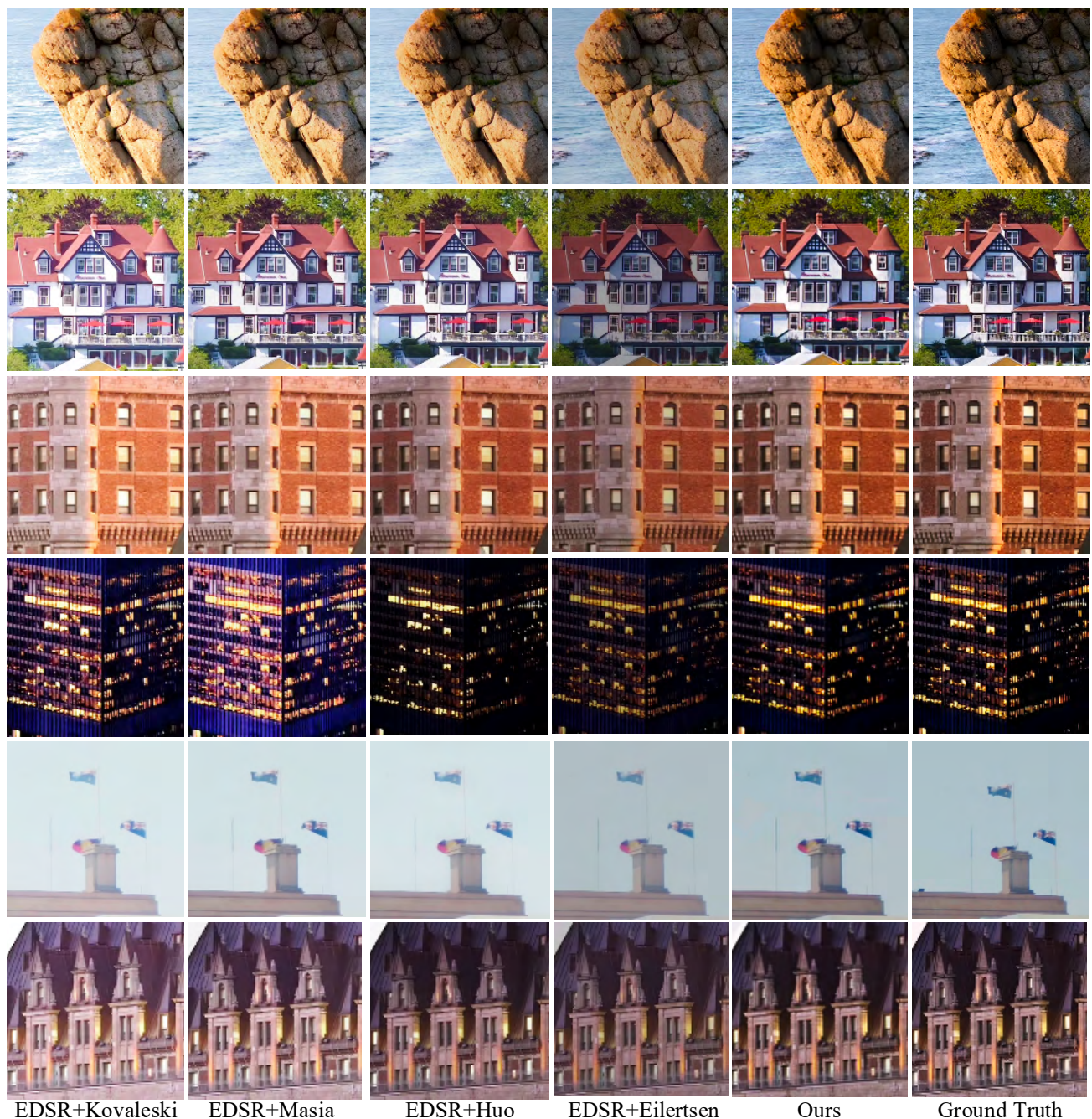

EDSR+Kovaleski     EDSR+Masia     EDSR+Huo     EDSR+Eilertsen     Ours     Ground Truth

Figure 2: Qualitative comparison against the cascade of EDSR [5] and ITM methods in [1-4].

### 1.3. Color artifacts in the CNN-based joint SR-ITM method

In Fig. 3, we provide more qualitative comparisons of Deep SR-ITM against the only CNN-based joint SR-ITM method by Kim *et al*. [6]. This multi-purpose CNN-based method [6] was trained from scratch on the same training data as ours. However, their method exhibits unnatural colors (oil-in-water type of artifact) and banding artifacts, especially in homogeneous regions such as the sky (columns 2 and 3) or water (column 1). In contrary, our method shows natural colors.

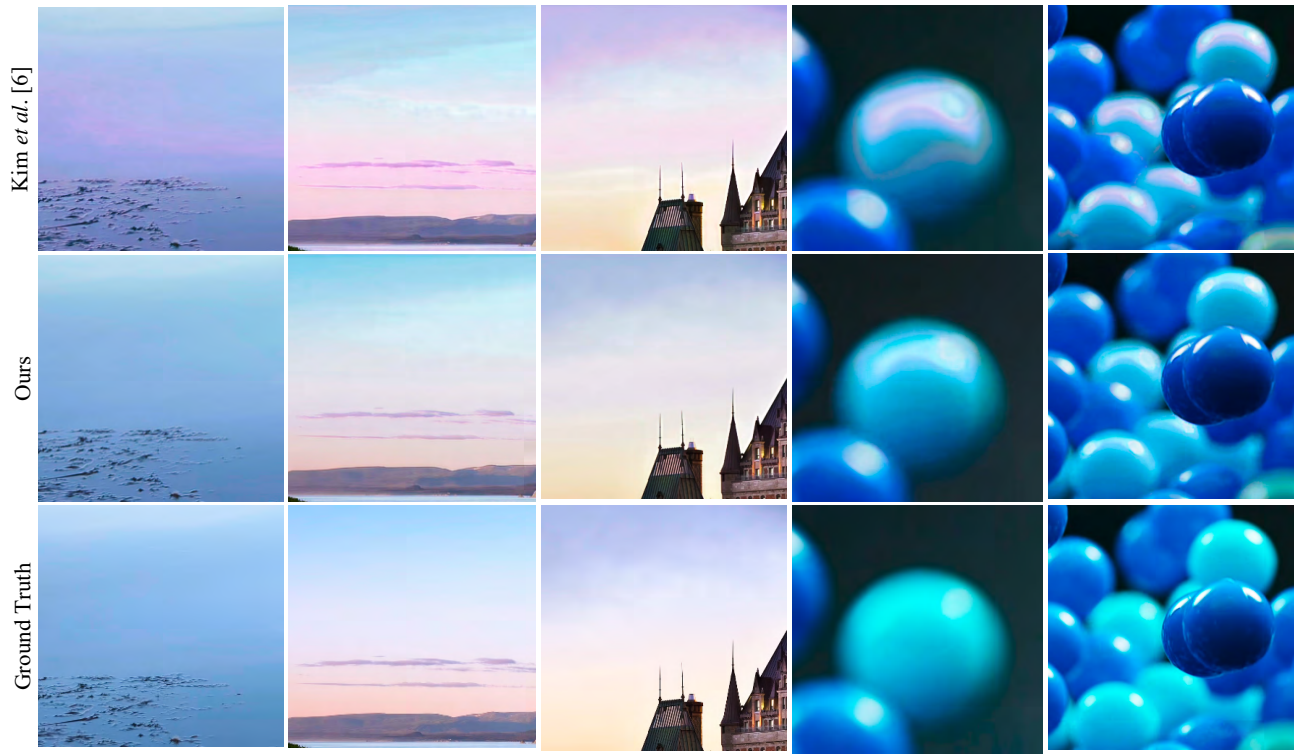

Figure 3: Qualitative comparison against Kim *et al*.'s method [6].

### 1.4. Comparison videos

We also provide four short comparison videos (two against the cascade of EDSR [5] and Eilertsen *et al.* [4], and two against the cascade of EDSR [5] and Huo *et al.* [3]) in 4K (3840×2160) resolution and HDR display format, along with this document as supplementary material. These videos were generated frame-by-frame from the input of 2K (1920×1080) SDR videos. We also provide the ground truth videos at the end of each clip. For Eilertsen *et al.*'s method [4], instead of resizing the image as in the authors' code, the test frame was cropped to the closest multiple of 32 (3840×2144), so that no pixel shift occurs after the 5-level U-Net. In the videos, the frames were zero-padded at the bottom to produce 4K resolution frames.

The provided videos are in the HDR10 format, and thus can be directly viewed with consumer HDR TVs. If such displays are unavailable, the reader is encouraged to view the videos using the MPC-HC[1] player with the madVR[2] renderer (on SDR displays). In this case, madVR must be selected in the Options-Playback-Output-Video tab in the MPC-HC player. We have heuristically found that this arrangement produces close results as viewing on HDR consumer displays. However, **note that the provided videos are not intended for directly viewing on SDR displays, and thus should not be judged based on any other settings than mentioned in this subsection.** We highly recommend the reader to view the videos, as they provide the comparisons in actual 4K resolution and HDR display format, which is difficult to express on paper. Please refer to the 'README.txt' file for details on the encoding specifications.

---

[1] https://mpc-hc.org/
[2] http://madvr.com/

## 2. Additional quantitative comparisons

We provide more quantitative comparisons against another SR baseline, LapSRN [7], cascaded with the ITM methods [1-3]. We use the official code released by the authors for [7]. Similar to Table 4 in our paper (comparison against EDSR [5] and [1-3]), Deep SR-ITM outperforms the cascade of previous methods in all measures and scale, except HDR-VDP for SF = 4. Also, similar range of values are observed with Table 4 of our paper, which implies that the influence of the SR method is minimal compared to the influence of the ITM method. This means that the compliance of overall color is a more important task in joint SR-ITM, than the reconstruction of high frequency details in the perspective of the final performance of the HR HDR prediction.

| Method | Scale | **PSNR (dB)** | **mPSNR (dB)** | **SSIM** | **MS-SSIM** | **HDR-VDP (Q)** |
|---|---|---|---|---|---|---|
| LapSRN [7]+Kovaleski et al. [1] | ×2 | 23.56 ± 0.96 | 25.45 ± 2.07 | 0.6487 ± 0.1551 | 0.9736 ± 0.0155 | 56.98 ± 3.61 |
| LapSRN [7]+Masia et al. [2] | ×2 | 24.70 ± 3.26 | 26.64 ± 5.16 | 0.7076 ± 0.1218 | 0.9675 ± 0.0330 | 59.94 ± 5.59 |
| LapSRN [7]+Huo et al. [3] | ×2 | 29.70 ± 2.54 | 31.74 ± 3.85 | 0.8930 ± 0.0703 | 0.9762 ± 0.0166 | 58.84 ± 6.39 |
| **Deep SR-ITM (Ours)** | ×2 | **35.58** ± 4.80 | **37.80** ± 5.83 | **0.9746** ± 0.0133 | **0.9839** ± 0.0097 | **61.39** ± 5.82 |
| LapSRN [7]+Kovaleski et al. [1] | ×4 | 23.40 ± 1.02 | 25.29 ± 2.06 | 0.6291 ± 0.1635 | 0.9654 ± 0.0151 | 56.16 ± 3.82 |
| LapSRN [7]+Masia et al. [2] | ×4 | 24.52 ± 3.18 | 26.45 ± 5.09 | 0.6953 ± 0.1174 | 0.9592 ± 0.0319 | **57.35** ± 5.22 |
| LapSRN [7]+Huo et al. [3] | ×4 | 28.74 ± 2.11 | 30.75 ± 3.36 | 0.8701 ± 0.0864 | 0.9678 ± 0.0157 | 55.19 ± 6.51 |
| **Deep SR-ITM (Ours)** | ×4 | **33.61** ± 4.32 | **35.73** ± 5.24 | **0.9561** ± 0.0259 | **0.9748** ± 0.0109 | 56.07 ± 6.83 |

Table 1: Qualititative performance comparison against the cascade of LapSRN [7] and ITM methods [1-3].

## Acknowledgement.

This work was supported by Institute for Information & communications Technology Promotion (IITP) grant funded by the Korea government (MSIT) (No. 2017-0-00419, Intelligent High Realistic Visual Processing for Smart Broadcasting Media).